\documentclass[twocolumn]{aastex63}

\usepackage{amsmath}
\usepackage{units}

\newcommand{\MESA}{{\tt MESA}}

\newcommand{\AESOPUS}{{\sc {\AE}SOPUS}}

\newcommand{\RCrB}{R\,CrB}

\newcommand{\Msun}{\ensuremath{\mathrm{M}_\odot}}
\newcommand{\Lsun}{\ensuremath{\mathrm{L}_\odot}}
\newcommand{\Msunyr}{\ensuremath{\mathrm{M}_\odot\,\mathrm{yr}^{-1}}}

\newcommand{\Mcore}{\ensuremath{\mathrm{M}_{\rm core}}}

\newcommand{\gcc}{\ensuremath{\mathrm{g\,cm^{-3}}}} 
\newcommand{\kms}{\ensuremath{\mathrm{km\,s^{-1}}}} 

\newcommand{\alphaMLT}{\ensuremath{\alpha_{\rm MLT}}}

\newcommand{\Teff}{\ensuremath{T_{\rm eff}}}

\newcommand{\logT}{\ensuremath{\log(T/\mathrm{K})}} 
\newcommand{\logRho}{\ensuremath{\log(\rho/\gcc)}} 

\newcommand{\nuclei}[2]{\ensuremath{\mathrm{^{#1}#2}}}

\newcommand{\hydrogen}[1][1]{\nuclei{#1}{H}}
\newcommand{\helium}[1][4]{\nuclei{#1}{He}}
\newcommand{\lithium}[1][7]{\nuclei{#1}{Li}}

\newcommand{\carbon}[1][12]{\nuclei{#1}{C}}
\newcommand{\nitrogen}[1][14]{\nuclei{#1}{N}}
\newcommand{\oxygen}[1][16]{\nuclei{#1}{O}}
\newcommand{\fluorine}[1][19]{\nuclei{#1}{F}}

\newcommand{\nickel}[1][58]{\nuclei{#1}{Ni}}

\begin{document}

\author[0000-0002-4870-8855]{Josiah Schwab}
\altaffiliation{Hubble Fellow}
\affiliation{Department of Astronomy and Astrophysics, University of California, Santa Cruz, CA 95064, USA}
\correspondingauthor{Josiah Schwab}
\email{jwschwab@ucsc.edu}

\title{Evolutionary models for R Coronae Borealis stars}

\begin{abstract}
  We use Modules for Experiments in Stellar Astrophysics (\MESA) to
  construct stellar evolution models that reach a hydrogen-deficient,
  carbon-rich giant phase like the R Coronae Borealis (\RCrB) stars.
  These models use opacities from OPAL and \AESOPUS\ that cover the
  conditions in the cool, H-deficient, CNO-enhanced envelopes of these
  stars.  We compare models that begin from homogeneous He stars with
  models constructed to reproduce the remnant structure shortly after
  the merger of a He and a CO white dwarf (WD).  We emphasize that
  models originating from merger scenarios have a thermal
  reconfiguration phase that can last up to $\approx \unit[1]{kyr}$
  post merger, suggesting some galactic objects should be in this
  phase.  We illustrate the important role of mass loss in setting the
  lifetimes of the \RCrB\ stars.  Using AGB-like mass loss
  prescriptions, models with CO WD primaries
  $\lesssim \unit[0.7]{\Msun}$ typically leave the \RCrB\ phase with
  total masses $\approx \unit[0.6-0.7]{\Msun}$, roughly independent of
  their total mass immediately post-merger.  This implies that the
  descendants of the \RCrB\ stars may have a relatively narrow
  range in mass and luminosity as extreme He stars and a relatively narrow
  range in mass as single WDs.
\end{abstract}


\section{Introduction}
\label{sec:intro}

The R Coronae Borealis (\RCrB) stars are hydrogen-deficient,
carbon-rich giant stars \citep{Clayton1996, Clayton2012}.  They are
notable for their high-amplitude photometric variability induced by
dust formation events in their atmospheres.  The objects are
closely related to the hydrogen-deficient carbon (HdC) stars and 
the extreme He (EHe) stars.  The origin of \RCrB\ stars has been
debated, through they are currently favored to be the outcome of
double white dwarf (WD) mergers \citep{Webbink1984}. 
Cementing this association requires stellar models
  that can follow the evolution from the conditions
  shortly after two WDs merge, to a subsequent \RCrB-like phase, and beyond.
  With these models in hand, one can then work to connect the properties
  of merging WDs with the properties of the observed stars,
  thereby constraining \RCrB\ progenitor systems and clarifying
  the physical processes that must operate during this evolution.

In this paper, we construct a set of evolutionary models for \RCrB\
stars using Modules for Experiments in Stellar Astrophysics (\MESA)
version r11701 \citep{Paxton2011, Paxton2013, Paxton2015, Paxton2018,
  Paxton2019}.
Section~\ref{sec:review} reviews the
substantial body of previous work constructing models of these stars.
As discussed in Section~\ref{sec:opacities}, we pay
special attention to the opacities, using the CO-enhanced OPAL
opacities \citep{Iglesias1993, Iglesias1996} supplemented at lower
temperatures by the CNO-enhanced \AESOPUS\ opacities
\citep{Marigo2009}.  In Section~\ref{sec:he-star}, we show models that
start as He stars and in Section~\ref{sec:schematic} compare with
models that start from conditions motivated by He WD + CO WD mergers.
In Section~\ref{sec:winds}, we illustrate the effects of the mass loss
prescription on the models and discuss its implications for \RCrB\ descendants.  In Section~\ref{sec:conclusion}, we
summarize and conclude.

\section{Review of Previous Work}
\label{sec:review}

The stellar structure of giants with helium-dominated envelopes and
degenerate carbon-oxygen cores was first explored by
\citet{Biermann1971} and \citet{Paczynski1971a}.
\citet{Schoenberner1977} also constructed models of such stars and evolved an $\unit[0.7]{\Msun}$ model through the \RCrB\ and EHe phases and onto the WD cooling track.
These studies
demonstrated that it was possible to construct He-shell-burning
stellar models with core masses $\approx \unit[0.7]{\Msun}$ that have
the luminosities and effective temperatures of the \RCrB\ stars.
Extending the models of \citet{Paczynski1971a}, \citet{Trimble1973}
found that helium stars in the mass range $\unit[0.8-2.0]{\Msun}$
become red giants.  While these studies could roughly explain the
structure of the \RCrB\ stars, they could not explain their formation.
In such models, the convective envelopes do not extend deeply enough
into the star in order to bring He-burning products (namely
\carbon[12]) to the surface.

\citet{Weiss1987b} constructed evolutionary models of \RCrB\ stars
that included the effects of the carbon and oxygen enhancements in the composition via specially prepared opacity tables.  This led to somewhat
lower effective temperatures than in previous work.  On the basis of
their luminosities and effective temperatures, these models suggested
that the \RCrB\ stars were constrained to be in the mass range
$0.8 \la M/\Msun \la 0.9$.

The understanding of the structure and properties of He shell-burning
stars was also advanced by the construction of equilibrium models.
\citet{Jeffery1988} explored the core mass-luminosity relationship for
these stars and \citet{Saio1988b} mapped out their luminosities and
effective temperatures as a function of the core and envelope masses.
Similarly, \citet{Iben1989} built families of
models with degenerate cores and He-burning shells, including a set of
models with pure He envelopes that they relate to the \RCrB\ stars.
Based on that understanding, \citet{Iben1990} studied the appearance of
the merger of He and CO WDs with time-dependent calculations, where the
merger is modeled as the rapid (super-Eddington) accretion of He on to
a CO WD.  This work argued that the thermal state of the core plays an important role, such that models evolved from He stars (with hot cores) appear different than models with cold cores from double WD mergers.

A more detailed comparison of the properties of similarly constructed
models of EHe (and \RCrB) stars was undertaken by \citet{Saio2002}.
Work in this vein has been continued by \citet{Zhang2014b} who
construct merger models in a similar fashion using \MESA.  They
consider models with both fast and slow accretion.  They find that
rapid accretion (their ``destroyed-disc'' models) can match the
observed carbon-rich abundances starting from lower He WD masses than
previous models.  \citet{Zhang2012b} also performed a similar study
evolving massive He+He WD mergers (total mass \unit[0.8]{\Msun}) which
may also provide a (rarer) channel for the formation of the \RCrB\ stars.

Broadly, these stellar evolution models confirmed the viability of the
double WD merger model for the origin of these systems.  However, the
detailed chemical abundances of the \RCrB\ stars provide another
strong constraint.  \citet{Asplund2000} used high-resolution spectra
and model atmospheres \citep{Asplund1997} to perform an abundance
analysis.  This work shows the \RCrB\ stars have enhanced CNO
abundances, generally with \mbox{C $>$ N $>$ O}.  Their nitrogen
abundances exceed that of CNO processed material at their metalicity.
This implies that there has been CNO-cycle processing of additional
material, presumably as a consequence of the merger.  Isotopically,
the CNO abundances are also peculiar.  The lack of \carbon[13] has
long been known \citep[e.g.,][]{Searle1961}.  More recently,
significant enhancements in \oxygen[18] were discovered
\citep{Clayton2007}, indicating the presence of conditions conducive
to $\alpha$-captures on \nitrogen[14].
These unusual surface abundances provide important clues to their
origins as they show a mix of H- and He-burning products.  As
discussed by \citet{Jeffery2011}, who use a simple model drawing on
abundances from detailed stellar evolution models to explore the
nucleosynthesis, the \RCrB\ abundances are consistent with the idea of
the merger of a He WD and CO WD, plus additional processing through H and
He burning that may occur during the merger.

Continuing progress in the modeling of the double WD merger process
itself has motivated work that uses the output from hydrodynamical
simulations to inform the conditions of material during and after the
merger.  For example, \citet{Longland2011, Longland2012} post-process
a merger simulation of an \unit[0.4]{\Msun} He WD and an
\unit[0.8]{\Msun} CO WD and characterize the nucleosynthesis that
occurs.
They find enhancements of \oxygen[18] and \fluorine[19] and demonstrate that
  \lithium[7] can be also be produced in hot WD merger events.
\citet{Staff2012} performed a series of hydrodynamic simulations of
merging double WD systems with constant mass (\unit[0.9]{\Msun}), but
varying mass ratios.  They found that lower mass ratio mergers
generally gave conditions more amenable to the production of
\oxygen[18], though the timescale over which these conditions persist
is uncertain.  Assuming that they continue for $\sim \unit[10^6]{s}$,
they found significant local production of \oxygen[18].  However, from
a global perspective, the large amount of \oxygen[16]-rich material
dredged-up during the merger prevents these calculations from matching
the \oxygen[18]/\oxygen[16] or the surface C/O ratio.  They suggested
this may imply that He-CO hybrid WDs, which have a thick He buffer
layer on their surfaces, are stronger candidates for the primary WD in
the merger.
A later series of hydrodynamic merger calculations including hybrid He-CO
WDs confirmed the idea that this He layer can prevent the dredge-up of
\oxygen[16]-rich material during the merger \citep{Staff2018}.

\citet{Menon2013} construct parameterized composition profiles that
schematically represent the outcome of the merger simulations of
\citet{Staff2012}.  They use abundances from more detailed stellar
models and use an amount of dredge-up less than that seen in the
hydrodynamic simulations.
They find that they are only able to reproduce the
surface abundances with a particular mixing profile.
The mixing must be neither too deep (or it will mix up additional \oxygen[16] and destroy \oxygen[18]) nor too shallow (or it will not mix up \oxygen[18]).
The mixing must also halt before the \RCrB\ star phase in order to preserve the
surface \nitrogen[14] abundance.
Qualitatively, one can link this to rotation and other processes in the merger.
An extension of this work to lower metallicity \citep{Menon2019}
finds \oxygen[16]/\oxygen[18] and \carbon[12]/\carbon[13] ratios
consistent with observed \RCrB s and
also with the possibility that these stars are the sources of some graphitic grains \citep{Karakas2015}.
The \RCrB\ models of \citet{Lauer2018} also use initial conditions
motivated by WD mergers.  Their models include the effects of
rotationally-induced mixing and use a 75-isotope nuclear network.
Focusing on the nucleosynthesis, they find significant \oxygen[18]
production and that their models are generally in agreement with the
observed abundances of a number of isotopes.

\section{Opacities}
\label{sec:opacities}

The outer layers of \RCrB\ stars are cool
$(\Teff \la\unit[10000]{K}$), hydrogen deficient, and carbon enhanced.
Such conditions are rarely encountered in standard stellar evolution
calculations and thus special attention should be paid to the adopted
microphysical inputs.  The evolutionary models of \citet{Weiss1987b}
spent significant effort to use a suitable equation of state
and opacities.

\citet{Weiss1987b} primarily studied models with two envelope
compositions.  The base metallicity of these models was $Z = 0.006$
\citep[with the solar abundance pattern from][]{Ross1976} plus carbon
and oxygen enhancements. Composition ``R1'' had $X_{\rm C} = 0.012$
while composition ``R2'' had $X_{\rm C} = 0.081$. Both compositions
had $X_{\rm O} = 0.01$.  Any remaining material was helium (so
$Y = 0.972$ and $Y = 0.903$ respectively).

\citet{Weiss1987b} used opacity tables from the Astrophysical
Opacity Library \citep{Huebner1977} with specially prepared extensions
to lower temperature by Huebner and Magee (1983, private communication
to A.~Weiss). These tables were later published in \citet{Weiss1990}:
the relevant tables are WKM20 (which corresponds to R1) and WKM21
(which corresponds to R2). The use of a uniform composition across the
parameter space minimizes the potential effects of interpolation
issues.

The default opacity tables in \MESA\ are not well-suited for
constructing evolutionary models of \RCrB\ stars.  \MESA\ does include
the OPAL radiative opacities for carbon and oxygen-rich mixtures
\citep{Iglesias1993, Iglesias1996}. These are referred to as OPAL
``Type 2'' tables and can be activated with the option
$\mathtt{use\_type2\_opacities}$.  When using these tables, the base
metallicity must also be selected using the control $\mathtt{Zbase}$.
We make use of the version of these tables calculated with the GS98
\citep{Grevesse1998} solar abundances.  The lower temperature boundary
of these OPAL tabulations is $\logT = 3.75$.

However, some of our \RCrB\ models extend to lower temperatures than
the OPAL opacities.  \MESA\ has not historically allowed for
low-temperature opacities that include separate carbon and oxygen
enhancements, though it now has the latent capability for this composition
dependence to exist throughout the parameter space.%
\footnote{A primary motivation is to allow for the inclusion of the
  effects of CNO-enhancements in hydrogen-rich material due to
  dredge-up on the AGB; however the infrastructure can equally well be
  applied to this hydrogen-deficient problem.
}
\MESA\ is usually forced to fall back to
opacity tabulations which assume a different composition.  The default
low temperature opacities are those of \citet{Ferguson2005} calculated
using a scaled-solar GS98 abundance pattern.  These can be evaluated
using either the base metalicity or the total metalicity, but in
either case, the assumed abundances do not reflect the composition of
the model.  This change in assumed composition means that when
blending between the OPAL tables and any of the included
low-temperature tables, there can be significant changes in opacity at
the location of the blend.

We used the web
interface\footnote{\url{http://stev.oapd.inaf.it/cgi-bin/aesopus}} to
the \AESOPUS\ opacities \citep{Marigo2009} to create a set of tables suitable for \RCrB\
stars. These tables and materials necessary to reproduce our work
are publicly available.%
\footnote{\url{https://doi.org/10.5281/zenodo.3386388}}
We generated tables with $X = 0$ and
base metallicities of $Z =$ 0.0006, 0.002, 0.006, and 0.02 (again using the GS98 solar abundance pattern).  We produced models with a wide range of CNO enhancement factors.  Carbon enhancement factors ($f_{\rm C}$) ranged from 1 to that needed to bring the carbon mass fraction to $\approx 10$ per cent.  Nitrogen enhancement factors ($f_{\rm N}$) range from 1 to 100 \citep[see Figure 8 in][]{Asplund2000}.  Carbon-to-oxygen mass ratios
($f_{\rm CO}$) were chosen to allow for oxygen enhancements while always remaining carbon-rich (so $\ga 0.4$).

\begin{figure}
  \centering
  \includegraphics[width=\columnwidth]{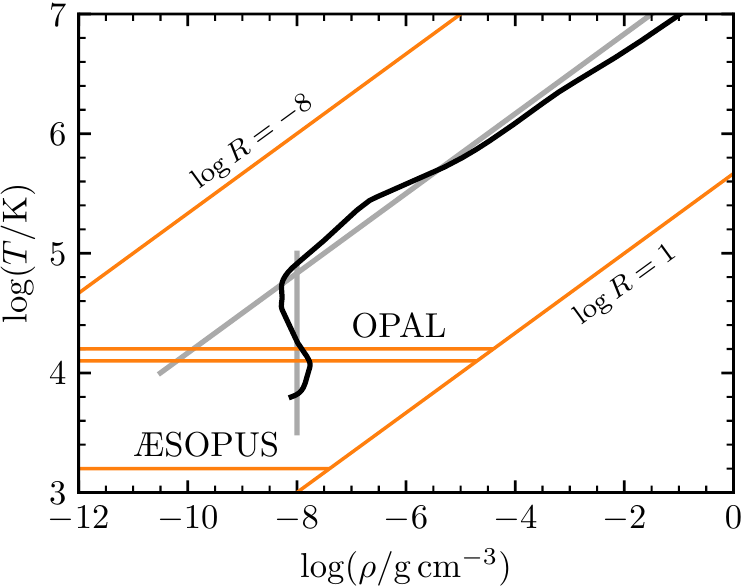}
  \caption{Opacity parameter space and table sources.  The solid black
    line shows a representative profile from a stellar model.  The
    grey lines indicate the slices of parameter space shown in
    Figures~\ref{fig:compare-kap-highT} and
    \ref{fig:compare-kap-lowT}.}
  \label{fig:kap-parameter-space}
\end{figure}

Figure~\ref{fig:kap-parameter-space} shows how these opacity sources
cover the $T-\rho$ parameter space.  We define the usual opacity parameter
$\log R = \logRho - 3 \logT + 18$.  For comparison, Figure 1 in
\citet{Paxton2011} illustrates default choices in \MESA\ and Figure 1
in \citet{Weiss1987b} shows a similar visualization of the inputs used
in that work.  The grey lines show slices of parameter space where we will compare the different opacities.

\begin{figure}
  \centering
  \includegraphics[width=\columnwidth]{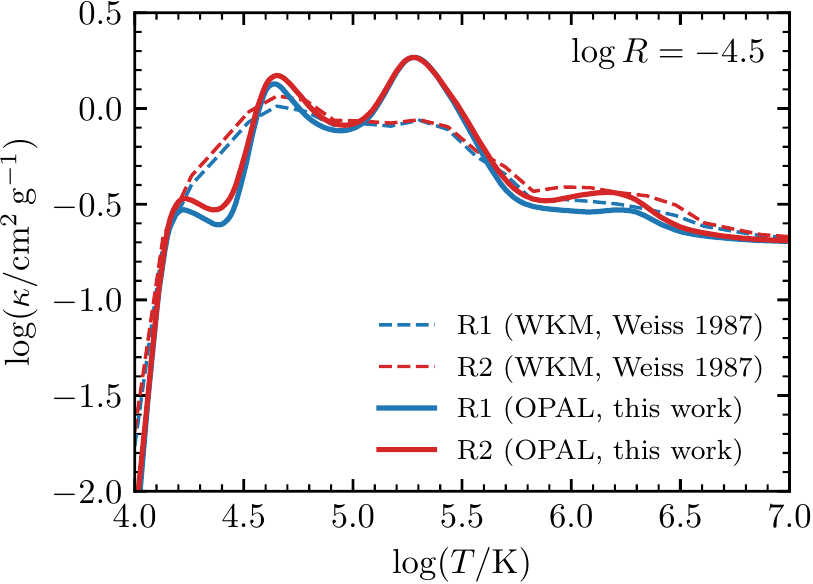}
  \caption{Comparison of opacities used in this work with those from
    \citet{Weiss1987b}.  The opacity is shown as a function of
    temperature at a constant value of $\log R$ (see diagonal grey
    line Figure~\ref{fig:kap-parameter-space}).  The more prominent Fe
    bump of more modern tables is particularly conspicuous.  These tables have $Z = 0.006$.}
  \label{fig:compare-kap-highT}
\end{figure}

Figure~\ref{fig:compare-kap-highT} compares the OPAL opacities used in this work with the earlier opacities used in \citet{Weiss1987b} at temperatures $\logT \ga 4.0$.  A more detailed comparison of these opacity sources and an explanation of the differences is presented in \citet{Iglesias1993}.  We show this primarily to remind the reader of the significant difference at the location of the Fe opacity bump around $\logT \approx 5.2$.

\begin{figure}
  \centering
  \includegraphics[width=\columnwidth]{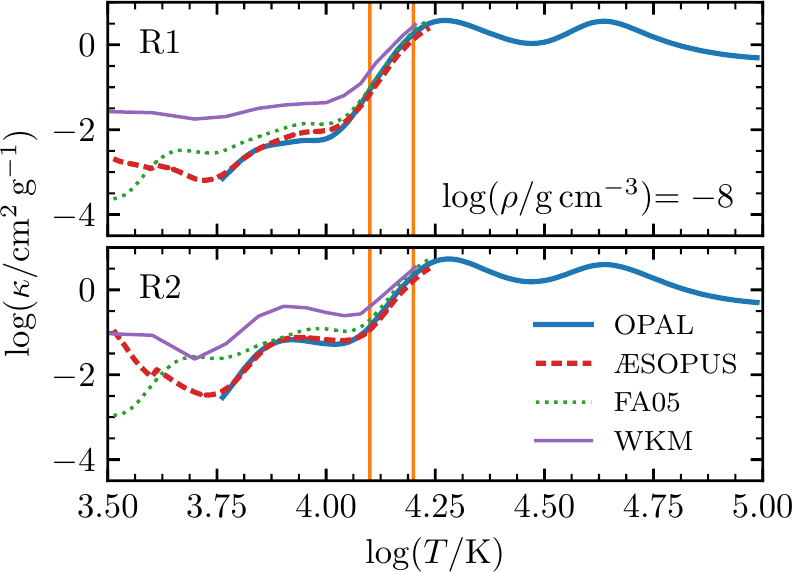}
  \caption{Comparison of opacities used in this work at lower
    temperatures.  The opacity is shown as a function of temperature
    at a constant value of $\rho$ (see vertical grey line
    Figure~\ref{fig:kap-parameter-space}).  The vertical orange lines
    indicate the location of the temperature blending region.  The
    OPAL and \AESOPUS\ opacities agree well in this temperature
    range.}
  \label{fig:compare-kap-lowT}
\end{figure}

Figure~\ref{fig:compare-kap-lowT} compares different opacity sources
at lower temperatures.  The OPAL and \AESOPUS\ tables agree well in
their overlap region.  We choose to locate the blend between the
tables at $4.1 \le \logT \le 4.2$, roughly in the middle of the overlap
region.  In this low temperature region, the WKM tables used in
\citet{Weiss1987b} report systematically higher opacities, though they
have similar shapes.  The FA05 \citep{Ferguson2005} opacities show
significant differences because they assume a scaled-solar
composition.  We show them because this is what would be used by
\MESA\ by default if we had not included more suitable tables.

\subsection{Atmosphere Boundary Conditions}
\label{sec:abc}

Early studies of \RCrB\ atmospheres were performed by
\citet{Myerscough1968} who included the effects of He
\citep{McDowell1966} and the ${\rm C}^{-}$ ion
\citep{Myerscough1964,Myerscough1966} as well as by
\citet{Schoenberner1975}.  One-dimensional, line-blanketed model
atmosphere calculations for these stars have been more recently
computed by \citet{Asplund1997} and \citet{Asplund2000}, but these are
not available in a form that allows for easy incorporation in a
stellar evolution code.  Therefore, our outer boundary conditions come
from atmosphere prescriptions that evaluate the Rosseland mean
opacities in the same way as in the rest of the stellar model.

We emphasize that the improved treatment of CNO-enhanced,
low-temperature opacities in this work should not be taken to indicate
that the absolute effective temperatures of our models are
particularly reliable.
The simplifications inherent in 1D mixing length theory
\citep{BohmVitense1958, CoxGiuli1968} -- whether in a stellar
evolution or model atmosphere code -- cannot reproduce the inherently
3D structure of convective envelopes, particularly in the
super-adiabatically stratified outer layers.
Promising future approaches include coupling of the results of 3D
hydrodynamical simulations with 1D stellar evolution calculations
\citep{Jorgensen2018, Jorgensen2019}.  However, details of where and how the boundary
condition is applied can still lead to effective temperature
ambiguities \citep[e.g.,][]{Choi2018a}.  We will illustrate the
sensitivity to the outer boundary conditions in Section~\ref{sec:obc}.

\subsection{Comparison with other work}

Recent work modeling \RCrB\ stars has generally not taken these
opacity effects into account.  \citet{Menon2013} primarily use \MESA\
with ``Type 1'' opacities (no C enhancement).  They do some test runs
using ``Type 2'' opacities, though these apply only at higher
temperatures (i.e., the OPAL tables).  They mention numerical
difficulties, which they attribute to the physical instability of
these envelopes.  Likewise, \citet{Zhang2014b} use the GS98 and FA05
tables.  They also experience numerical difficulties that lead to
small fluctuations in the luminosity.  Similar fluctuations are
apparent in the HR diagrams in \citet{Lauer2018}.  For the most part
the models in the current work evolve smoothly when moving to the blue
suggesting that some of these difficulties may have reflected a poor
blend between the low and high temperature opacity tables.  The
exception is the models with the highest metallicity and highest
luminosities, which are locally super-Eddington at the Fe opacity
bump; such models continue to pose numerical challenges.

\section{Helium Star Models}
\label{sec:he-star}

We first construct \RCrB\ models via a modified evolution of helium
stars.  These are closely related to the homogeneous models of
\citet{Weiss1987b}.  We begin the evolution from homogeneous models on
the He ZAMS. We do not activate the predictive mixing
capabilities described in \citet{Paxton2018}.  This may result in an
underestimate of size of the convective core during the He MS, but we
are not interested in the properties of the model during core burning.
We let these models move into the shell burning phase and allow the CO
cores to grow.  When the envelope mass shrinks to 0.47 \Msun, we stop
the models and instantaneously change the envelope composition.  We
then resume the evolution and continue until the model reaches
$\Teff = \unit[30]{kK}$ while moving to the blue.  Only this latter
portion of the evolution is shown in the paper.
We do not include the effects of mass loss, so each model has a
constant total mass.  The primary goal of this section is to compare
with past work and illustrate the sensitivity to various modeling
assumptions.

\subsection{Comparison with \citet{Weiss1987b}}

\begin{figure}
  \centering
  \includegraphics[width=\columnwidth]{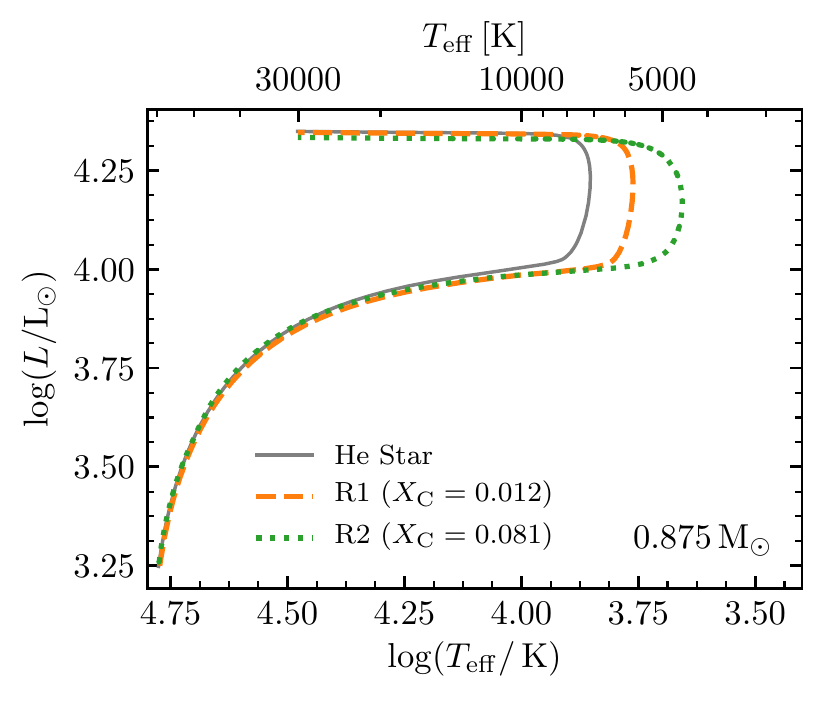}
  \caption{Evolution of He star models with varying envelope compositions.   This reproduces the results of earlier work \citep[cf.~Figure 3b in][]{Weiss1987b}.}
  \label{fig:weiss-HR}
\end{figure}

As a first illustration of these models, Figure~\ref{fig:weiss-HR}
shows 3 tracks in the HR diagram.  These correspond to models of
$Z = 0.006$ He stars with an unmodified envelope composition and with
compositions R1 and R2.  These results are in general agreement
with those in \citet{Weiss1987b}.  The models have similar
luminosities during their redward and blueward evolution,
with the \MESA\ models being slightly $(\lesssim 0.1$ dex) warmer
at their coolest point.

\subsection{Sensitivity to Outer Boundary Conditions}
\label{sec:obc}

The effective temperatures of our models depend on the outer boundary
condition, the mixing length parameter, and the opacities used in the
model (see Section \ref{sec:abc} for more discussion).  Here we
illustrate the shifts caused by different options.  The illustrative
model in used this subsection has $M = \unit[0.85]{\Msun}$,
$Z = 0.006$, and envelope composition R1.

\begin{figure}
  \centering
  \includegraphics[width=\columnwidth]{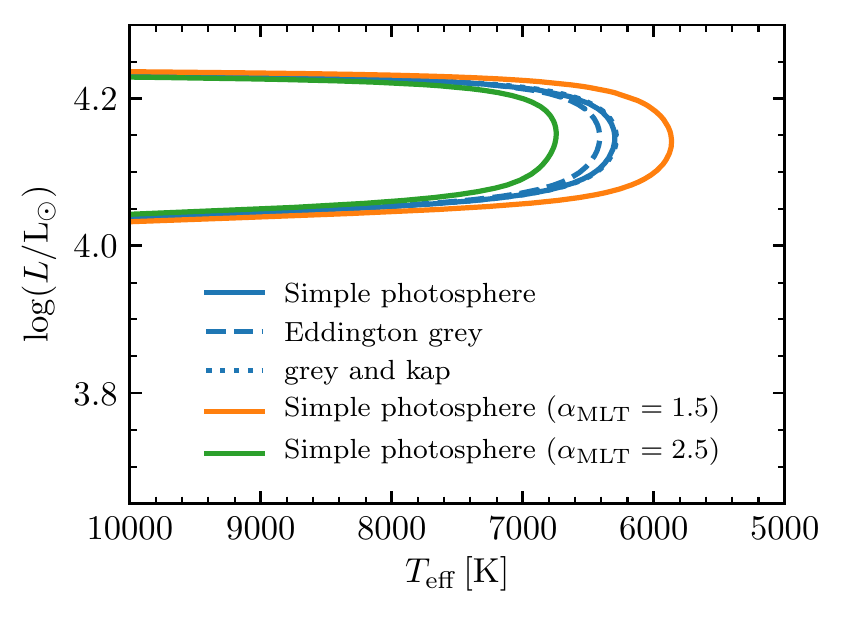}
  \caption{Effect of outer boundary condition and mixing length parameter on model effective temperatures.}
  \label{fig:BCs}
\end{figure}

Throughout, we default to using the \MESA\ ``simple photosphere''
option, which uses a simple, constant grey opacity solution to the
radiative diffusion equation \citep[see Section 5.3
in][]{Paxton2011}.
Figure~\ref{fig:BCs}
illustrates the results of other choices.  We show the results with
the ``grey and kap'' option, which is like the simple photosphere
option, but with an additional iterative step to ensure that the
pressure, temperature, and opacity at the surface are consistent.  We
also show the ``Eddington grey'' option, which integrates the $T-\tau$
relation of \citet{Eddington1926}.  These options all agree at level
of $\approx \unit[100]{K}$.

Figure~\ref{fig:BCs} also shows the effect of varying the mixing
length parameter \alphaMLT.  (The default value is
$\alphaMLT = 2$.)  This shift is at the level of
$\approx \unit[1000]{K}$.  This reflects the fact that as these He
stars reach a substantial fraction of the Eddington luminosity,
convection is becoming inefficient.  This leads to density inversions
in the outer layers and the radius to be particularly sensitive to
\alphaMLT\ \citep[e.g.,][]{Joss1973, Sanyal2015}.

\subsection{Sensitivity to Composition and Opacities}
\label{sec:composition}

Our fiducial composition has a base metallicity of $Z = 0.006$.  We
adopt the CNO abundances of the majority \RCrB\ population from
\citet{Asplund2000}.  Their Table 6 gives, relative to the GS98 solar
abundance pattern, a nitrogen enhancement factor
$\log(f_{\rm N}) = 1.7$ and an oxygen enhancement factor
$\log(f_{\rm O}) = 0.4$.  We split the oxygen equally by mass between
\oxygen[16] and \oxygen[18], though in the opacities there is no
distinction between these isotopes.

The carbon abundances are somewhat less clear due to difficulty in
modeling the \ion{C}{1} lines \citep[i.e., the carbon
problem;][]{Gustafsson1996}. \citet{Asplund2000} explore C/He number
ratios in the range 0.1 to 10 per cent, finding ratios $\gtrsim 3\%$
are ruled out for the majority population.  The EHe stars appear to
have somewhat lower values with C/He $\approx 1\%$ \citep[see Section
3.3.3 of][and references therein]{Asplund2000}.
Our fiducial carbon enhancement factor is $\log(f_{\rm C}) = 1.2$,
corresponding to a carbon mass fraction of $\approx 0.04$, and hence a
number ratio of $\approx 1\%$.

\begin{figure}
  \centering
  \includegraphics[width=\columnwidth]{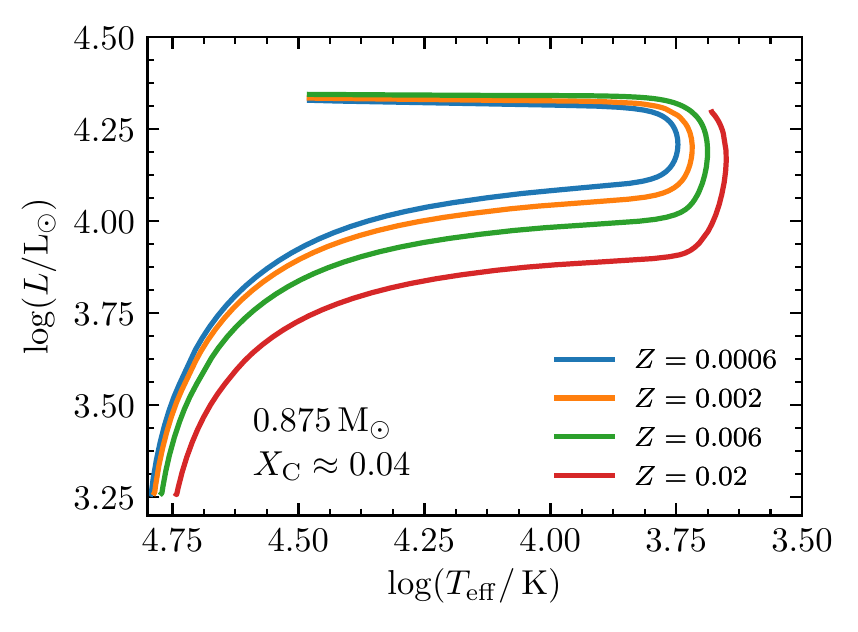}
  \caption{Effect of varying base metallicity at fixed mass and
    envelope composition. The highest metallicity model experiences
    numerical issues in the outer layers before evolving back to the
    blue.}
  \label{fig:metallicity}
\end{figure}

Figure~\ref{fig:metallicity} illustrates the effect of varying the
metallicity at a fixed mass of $\unit[0.875]{\Msun}$.  All models have
the exact same composition in terms of He and CNO elements, but the
base metallicity of the opacity tables is varied.  This illustrates
the influence of the Fe-bump on the radius of the envelope.  The
$Z = 0.02$ model experiences numerical problems.  As the opacity
and/or luminosity rise, convection becomes increasing inefficient,
requiring an increasingly super-adiabatic temperature gradient to
transport the energy.  As a result a steep entropy gradient develops
at the base of the convection zone.  Resolving this feature and its
Lagrangian movement (as the envelope mass changes due to the He shell
burning at the base) requires short timesteps and can also trigger
numerical convergence issues.  (See the \MESA\ He star models of
\citealt{Hall2018} for another example of this issue).  These could be
circumvented by artificially increasing the efficiency of convection
\citep[e.g., using the MLT++ prescription;][]{Paxton2013}, though that
does not necessarily produce reliable radius estimates.

\begin{figure}
  \centering
  \includegraphics[width=\columnwidth]{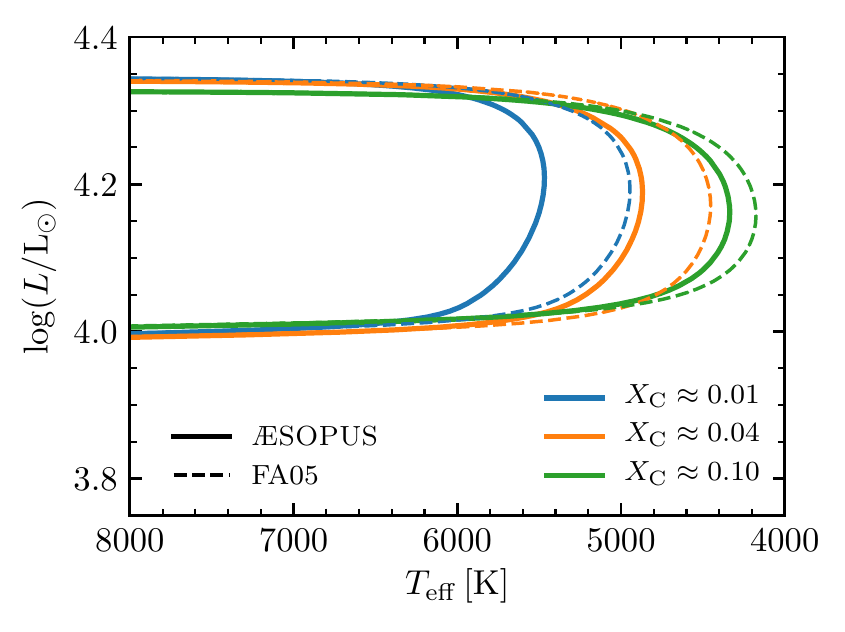}
  \caption{Change in effective temperature when varying surface carbon
    fraction at a fixed base metallicity of $Z = 0.006$.  The solid
    lines show the results using the CNO-enhanced \AESOPUS\ opacities.
    The thin dashed lines show the calculations, but using the
    scaled-solar FA05 opacities.}
  \label{fig:carbon}
\end{figure}

Figure~\ref{fig:carbon} shows the effect of changing the envelope
carbon fraction at a fixed base metallicity of $Z = 0.006$.  This is a
similar exercise as Figure~\ref{fig:weiss-HR}, except that we also
show models when the scaled-solar FA05 opacities are used instead of
the CNO-enhanced \AESOPUS\ opacities.  This illustrates the level of
difference between previous models using versions of \MESA\ that only
had scaled-solar low-temperature opacity options available.  Models
are generally $\approx \unit[500]{K}$ warmer.

\section{Merger Models}
\label{sec:schematic}

The He star models presented in Section~\ref{sec:he-star} are a simple
way to construct an \RCrB-like object.  However, their configuration
may be somewhat different than that realized in a WD merger.  In
particular, \citet{Iben1990} emphasizes the importance of the thermal
state of the core.
In the WD merger case, one may expect the CO WD
that forms the core to be more degenerate than the CO core created within the
He star at the conclusion of He core burning.%
\footnote{This is the reason that WD mergers are able to form He giants with masses below the mass where a single He star would become a giant.}
Therefore, in this
section, we construct models more directly motivated by the
post-merger configuration of a double WD merger.
We initialize \MESA\ models with cold CO cores and high entropy He-rich envelopes and discuss their evolution.

\subsection{Detailed Merger Model}
\label{sec:ZP4}

\citet{Schwab2012} performed multi-dimensional simulations of the
viscous phase of the merger (using an $\alpha$-viscosity
prescription), beginning from the output of SPH simulations of the
dynamical phase of the merger \citep{Dan2011} and covering the
hours-long phase where the rotationally-supported disc transitions
into a spherical, thermally-supported envelope \citep{Shen2012}.  We
focus on the model ZP4, which is the merger of an $\unit[0.3]{\Msun}$
He WD with an $\unit[0.6]{\Msun}$ CO WD.  A small amount of material
is lost during the merger and the remnant has a total mass
$\unit[0.88]{\Msun}$.

\begin{figure}
  \centering
  \includegraphics[width=\columnwidth]{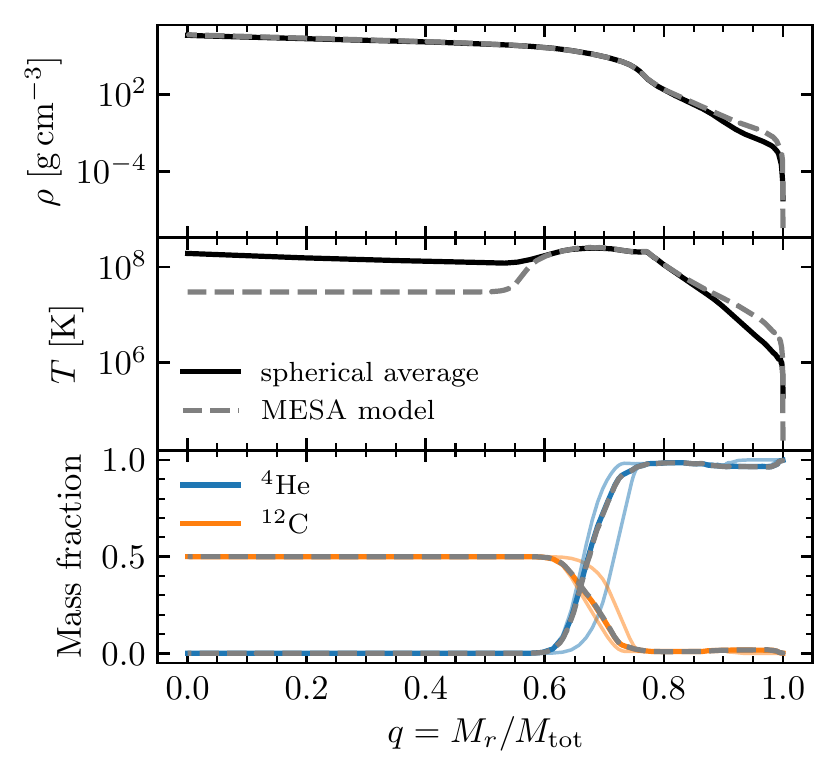}
  \caption{Comparison of density, temperature, and composition
    profiles between spherical averages taken at the end of the
    hydrodynamic calculations of the viscous disc phase \citep[model
    ZP4 from][]{Schwab2012} and the initial \MESA\ model.  In the
    lower panel, the thin lines indicate the composition profiles
    taken along polar and equatorial slices.}
  \label{fig:ZP4}
\end{figure}

In the same manner as \citet{Schwab2016b}, we take the final output of
the \citet{Schwab2012} calculation, spherically-average the entropy
and composition profiles, and then generate \MESA\ models that
approximately match these profiles.  The initial state of this model
is shown in Figure~\ref{fig:ZP4}.  We do not include the effects of
rotation, though the outer layers of the model do still have some
rotational support, which is the primary source of the disagreement
between the spherical averages and the \MESA\ model at
$q \gtrsim 0.9$.  Additionally, in the cool outer layers
$(\logT \lesssim 7)$, there is a mismatch in the equation of state
between the hydrodynamics calculations and \MESA.  The hydrodynamics
calculations use the Helmholtz EOS \citep{Timmes2000b} assuming full
ionization, whereas the \MESA\ EOS accounts for ionization via the use
of OPAL \citep{Rogers1996} and PTEH \citep{Pols1995}.  Therefore, the
detailed thermal structure of the very outer parts of the \MESA\ model is
unlikely to be particularly reliable.  However, the region around the temperature peak where most of the thermal energy resides is a good match.

As noted in \citet{Schwab2012}, while the overall remnant in this case
is spherical, the composition profile varies with polar angle.  We
illustrate this with the thin lines in the lower panel of Figure~\ref{fig:ZP4}, which show the
composition profiles along polar and equatorial slices.  The spherical
averaging necessary to construct the 1D \MESA\ models leads to some
unavoidable chemical mixing in the interface region.

In the models of \citet{Dan2011} the He WDs are composed of \helium[4]
(and the CO WDs have equal mass fractions of \carbon[12] and
\oxygen[16]).  The hydrodynamic evolution uses a minimal (7 isotope)
nuclear network.  Likewise, the calculations in \citet{Schwab2012} use
a small network to track the energy release from He burning.  This
means that these calculations contain little information about the
nucleosynthesis in the merger and its immediate aftermath.  In
particular, directly confronting the observed surface abundances of
\RCrB\ stars requires a more detailed chemical model of the He WD and
a larger nuclear network to follow the key chemical conversions (e.g.,
CNO cycling from \hydrogen[1], \lithium[7] from \helium[3],
\oxygen[18] from \nitrogen[14]).  Work such as \citet{Lauer2018}
performs \MESA\ calculations using a larger nuclear network and other
work post-processes with yet larger ones
\citep{Longland2012, Menon2013}.  Here, given the limitations of our
modeling approach, we choose not to focus on the nucleosynthesis.

\subsection{Schematic Merger Models}
\label{sec:grid}

We also generate some models that are not based directly on
the result of merger calculations, but instead schematically reproduce
configurations with a cold core and thermally-supported envelope.
Given the uncertainties in the merger modeling, these help to provide
a sense of which evolutionary features are robust.  Because they are
simple to construct, they also allow for rapid exploration of the
parameter space.

We pick a CO core mass and a He envelope mass and assume a sharp
core-envelope boundary.  For simplicity, all models have the same core
temperature ($T = \unit[3\times10^7]{K}$).  We initialize the envelope
with a constant specific entropy.  By default, we assume
$s = \unit[10^9]{\rm erg\,g^{-1}\,K^{-1}}$.  This approach is similar
to that applied in \citet{Shen2012} and the relaxation procedure used
was adapted from publicly-available code \citep{ShenSS}.  The core
composition is equal mass fractions of \carbon[12] and \oxygen[16].
The envelope initially has the same fiducial composition described in
Section~\ref{sec:composition}.

\begin{figure}
  \centering
  \includegraphics[width=\columnwidth]{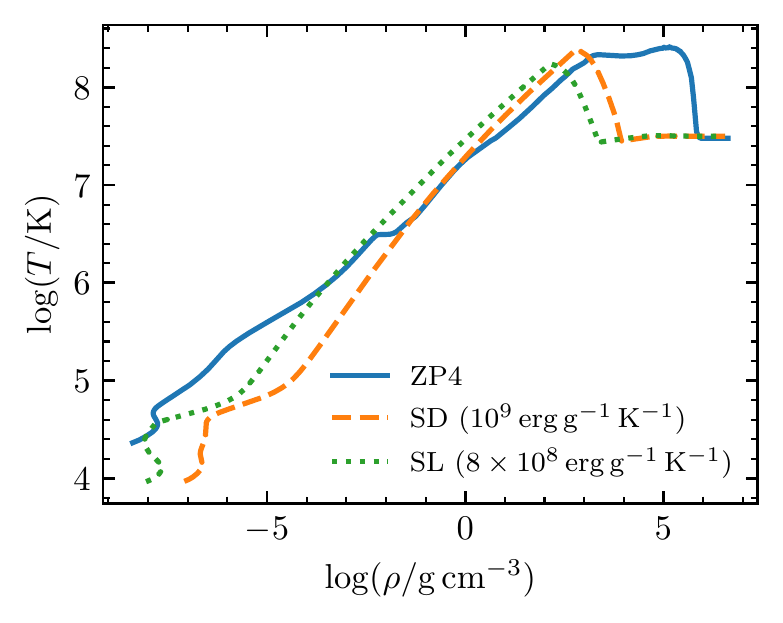}
  \caption{Initial temperature-density profiles of merger models.
    Model ZP4 is the detailed merger model described in
    Section~\ref{sec:ZP4}.  Model SD is the schematic merger model
    described in Section~\ref{sec:grid} for a $\unit[0.3]{\Msun}$ He
    WD and $\unit[0.6]{\Msun}$ CO WD.  Model SL is the same as model
    SD, but with a lower specific envelope entropy of
    $s = \unit[8\times10^8]{\rm erg\,g^{-1}\,K^{-1}}$.}
  \label{fig:TRho-schematic}
\end{figure}

Figure~\ref{fig:TRho-schematic} shows the initial temperature and
density profile for models with different envelope entropies.
Compared with the model ZP4 from Section~\ref{sec:ZP4}, the peak
temperature is similar, though generally at lower density in the
schematic models (SD and SL).

\subsection{Comparative Evolution}

The primary qualitative difference between these models and the He
star models of Section~\ref{sec:he-star} lies in the early time
evolution.  The merged remnant is far from thermal equilibrium.  The
base of the envelope must first set up the steady He-burning shell and
then the rest of the envelope adjusts to this input luminosity.  The
post merger configuration is generally compact, and so initially the
thermal energy deposited in the merger and released from nuclear
burning goes into work leading to expansion.  As a result, the objects
begin at relatively lower luminosities that then increase as they move
towards the steady He shell burning configuration.

\begin{figure}
  \centering
  \includegraphics[width=\columnwidth]{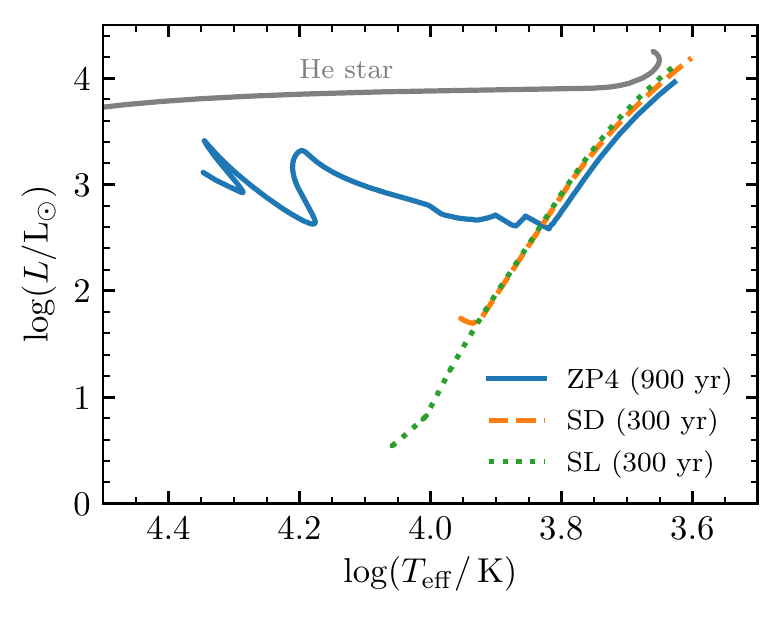}
  \caption{Evolution in the HR diagram shortly after merger.
    Indicated times are the duration from the start of the calculation
    until the surface luminosity and the luminosity from He burning
    are first equal.  The grey line shows the qualitatively different
    evolutionary track from a homogeneous He star model.}
  \label{fig:early-HR}
\end{figure}

Figure~\ref{fig:early-HR} shows the early post-merger evolution of the
models shown in Figure~\ref{fig:TRho-schematic}.  As a consequence of
the differences in the initial models, the behavior is different at
first.  However, by the time they reach $\log(L/\Lsun) \gtrsim 2.5$
all models are similar.  The tracks continue until
$L \approx L_{\rm He}$ (surface luminosity is approximately nuclear luminosity from He burning).  The
amount of time elapsed in the simulation is indicated in the legend,
ranging from $\unit[300]{yr}$ to $\unit[900]{yr}$.  We suspect that
the longer time corresponding the ZP4 model is more physically
realistic.  In the SPH merger models of \citet{Dan2014}, a $q = 0.5$
merger has only about half its mass in the cold core (with the
remainder roughly equally distributed between a disk and hot
envelope).  In contrast, our simple schematic merger models have the
entire mass of the primary (i.e., two-thirds of the total mass) in
their cold cores.  The schematic merger models thus represent an
extreme limit and likely have systematically shorter thermal times
from the temperature peak to the surface than models based on merger
calculations where the outer layers of the primary have been strongly
heated during the merger.

The existence of this thermal reconfiguration phase is one of the
hallmarks of double WD mergers.
It is not present in the early evolution of models coming from
homogeneous He stars \citep[e.g.,
Section~\ref{sec:he-star};][]{Weiss1987b, Menon2013}.
The \citet{Lauer2018} models, constructed similarly to the schematic
models, also show this early brightening phase.  In the
\citet{Zhang2014b} destroyed disc models this segment of the evolution
lasts $\approx \unit[500]{yr}$ (see their Figure 18).

Given \RCrB\ lifetimes $\sim \unit[5\times10^4]{yr}$, the relative
duration of this phase is $\sim 1 \%$.  The galactic \RCrB\ stars
number in the hundreds \citep[e.g.,][]{Tisserand2018b}.
The relative duration suggests the presence of at least
  a few objects in this pre-\RCrB\ phase.  These objects
  may not yet exhibit the complex photometric behavior
  characteristic of the \RCrB\ stars, and so may not be classified as such.
These pre-\RCrB\ objects
will be gradually growing brighter and cooler on timescales comparable
to the baseline covered by historical photometric catalogs.  They also
have likely not yet shed significant quantities of dusty material and
so should have different circumstellar environments from objects that
are settled in the \RCrB\ phase.

\section{Effects of Mass Loss}
\label{sec:winds}

Mass loss plays an important role in \RCrB\ evolution.  Dusty shells, with dust masses up to $\unit[10^{-3}]{\Msun}$, 
exist in the surroundings of \RCrB\ stars and are interpreted to have
formed during the \RCrB\ phase \citep{Montiel2015, Montiel2018}.  The
terminal velocities of \RCrB\ star winds are
$v_{\infty} \approx \unit[300]{\kms}$ \citep{Clayton2003,
  Clayton2013}.  Note that this is $\gtrsim 10$ times greater than
typical carbon-rich AGB wind velocities
\citep[e.g.,][]{Groenewegen1998}.
Recent evolutionary models of \RCrB\ stars have adopted the AGB wind
prescription of \citet{Bloecker1995a}, with efficiency factors
$\eta \approx 0.02$ \citep{Menon2013, Zhang2014b, Lauer2018}.
Typically mass loss rates are then
$\sim \unit[10^{-6} - 10^{-5}]{\Msunyr}$.

The application of existing mass loss prescriptions to the
  \RCrB\ stars comes with significant caveats.  The \RCrB\ stars have
  different envelope compositions (H-deficient, C-rich) than AGB stars
  (normal H, often O-rich).  Their pulsation periods are also shorter
  ($\lesssim \unit[100]{d}$) than those of the long-period variables
  that motivate prescriptions like that of \citet{Bloecker1995a}.
  Even if AGB mass loss were a solved problem, the solutions would
  not be directly applicable to the \RCrB\ stars.
  Nonetheless, these existing prescriptions provide a convenient
  initial guess for exploration.

The merger creates a He envelope of $\approx \unit[0.3]{\Msun}$ on a CO core.
Then, the lifetime of the \RCrB\ phase is given by the timescale to
exhaust this envelope as material is added to the CO core through
He-burning $(\dot{M}_{\rm He})$ and removed from the star through
winds $(\dot{M}_{\rm wind})$.
First, note $\dot{M}_{\rm He} = L / Q_{\rm He}$ and $L$ is a strong
function of the CO core mass $M_{\rm core}$.  For example,
$d\ln L/d\ln \Mcore \approx 5$ for $\Mcore \approx \unit[0.6]{\Msun}$
\citep{Jeffery1988}.
Second, note that $\dot{M}_{\rm wind}$ also likely increases with
increasing $L$.  Thus, the envelope is exhausted more rapidly by both
processes as the core grows, meaning that more time is spent at lower
core masses in any given model.

If the wind mass loss rates grow strongly with $L$---the
\citet{Bloecker1995a} prescription has
$\dot{M} \propto L^{3.7}$---then this effectively creates a critical
core mass defined by the point where
$\dot{M}_{\rm wind} \approx \dot{M}_{\rm He}$.  For initial core
masses below this critical mass, the core can grow.  But once it
reaches the critical mass (or if the initial core mass is above it),
the large luminosity causes the envelope to be rapidly shed.
Therefore, though there can be a significant range in the initial
total mass of the remnant, this may be erased by mass loss.

The wind velocity implies an upper limit to the mass loss rate of
\begin{equation}
  \dot{M}_{\rm max} \approx \frac{2L}{v_{\infty}^2} \sim
  \unit[10^{-3}]{\Msunyr}
  \left(\frac{L}{\unit[10^4]{\Lsun}}\right)
  \left(\frac{v_{\infty}}{\unit[300]{\kms}}\right)^{-2}~,
  \label{eq:maxmdot}
\end{equation}
on energetic grounds.
Comparing the wind specific kinetic energy to the specific energy of
He-burning, we see
\begin{equation}
  v_{\infty}^2 \approx \unit[10^{15}]{erg\,g^{-1}} \ll Q_{\rm He} \approx \unit[7\times10^{17}]{erg\,g^{-1}} ~.
\end{equation}
The ad hoc assumption that $\dot{M} = f \dot{M}_{\rm max}$ (with a constant value of $f$) provides a simple expression that we will use to draw a contrast with the \citet{Bloecker1995a} prescription and its much steeper $L$ dependence.  Under this assumption, the growth of the core is
\begin{equation}
\Delta M = M_{\rm He} \left(1 + \frac{2fQ_{\rm He}}{v_{\infty}^2}\right)^{-1}~.
\end{equation}
Then for $f \gtrsim v_\infty^2 / (2 Q_{\rm He}) \sim 10^{-3}$, only a
small amount of the He will be added to the CO core and the rest will
be lost from the star.

\begin{figure}
  \centering
  \includegraphics[width=\columnwidth]{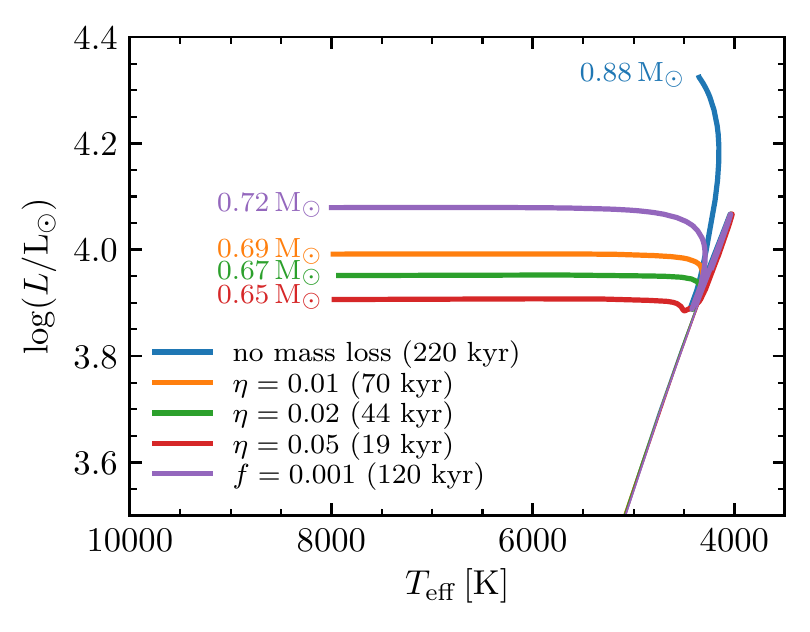}
  \caption{Evolutionary dependence on mass loss rate.  Total masses
    are indicated at the end of each track.  Lifetimes are indicated
    in the legend.  The thin portion of the track is the early thermal
    adjustment phase (shown in more detail in
    Figure~\ref{fig:early-HR}) and is not included in the stated
    lifetime.  Models labeled with $\eta$ values use a
    \citet{Bloecker1995a} wind with that scaling factor.  The model
    labeled with an $f$ value uses a wind that with that constant
    fraction of the maximum given by Equation~\eqref{eq:maxmdot}.}
  \label{fig:lifetime}
\end{figure}

Figure~\ref{fig:lifetime} shows initial model ZP4 evolved with the
\citet{Bloecker1995a} wind prescription in \MESA\ using a range of
efficiency factors $\eta$.  Once the He envelope shrinks to
$\approx \unit[0.05]{\Msun}$ the models begin to evolve to the blue.
The case without mass loss reaches a near-Eddington luminosity and we
have halted this model at a CO core mass $\approx \unit[0.8]{\Msun}$,
when the timestep becomes severely limited (see
Section~\ref{sec:composition} for a discussion of this issue).
For the models including mass loss, the total mass does not have a
strong dependence on $\eta$, and they evolve to the blue with masses
$\approx \unit[0.65 - 0.70]{\Msun}$.  However, the lifetime is more
sensitive, with the factor of 5 variation in $\eta$ giving a factor of
$\approx 3$ change in lifetime.  These lifetimes and masses are
consistent with past work.  The models of \citet{Zhang2014b} typically
spend $\unit[50–70]{kyr}$ in the \RCrB\ phase; the models of
\citet{Lauer2018} spend $\sim \unit[10^5]{yr}$.  Similarly, both works
show the total masses being reduced to $\approx \unit[0.7]{\Msun}$.
The model in Figure~\ref{fig:lifetime} with $f = 0.001$ illustrates a
different prescription giving a similar final mass but significantly
different lifetime.

Figure~\ref{fig:grid-masses} shows the total mass of each of a grid of
schematic merger models (all with CO primaries less massive than
$\unit[0.7]{\Msun}$) at a time after they have evolved to the blue
(reached $\Teff = \unit[10^4]{K}$).  We compare two mass loss
prescriptions: the top plot (a) uses the \citet{Bloecker1995a} wind
with $\eta = 0.02$; the bottom plot (b) uses
$\dot{M} = f \dot{M}_{\rm max}$ with $f = 0.001$.  The systems have a
range of initial total masses $\unit[0.7-1.05]{\Msun}$. In plot (a),
all leave the \RCrB\ phase with total masses in the range
$\approx \unit[0.6-0.7]{\Msun}$.  In plot (b), the systems have masses
$\approx \unit[0.65-0.8]{\Msun}$, more massive on average and with a
slightly wider mass range than those in plot (a).  This is as expected
given the mass loss rates in plot (b) increase less rapidly as $L$
grows.  Though the exact numbers will depend on the implemented
prescription, this illustrates the inevitable effect of mass loss to
concentrate remnants toward lower total masses.

\subsection{Implications for \RCrB\ descendants}

\begin{figure}
  \centering
  \fig{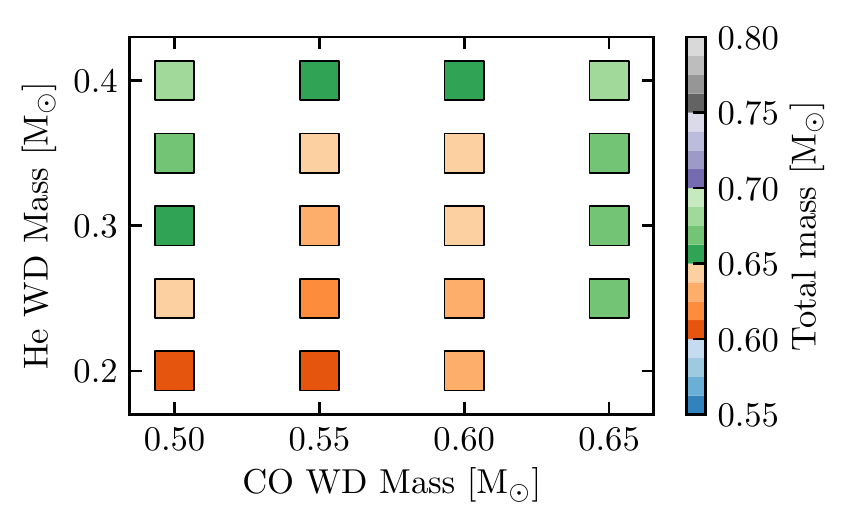}{\columnwidth}{(a) \citet{Bloecker1995a} wind using $\eta = 0.02$.}
  \fig{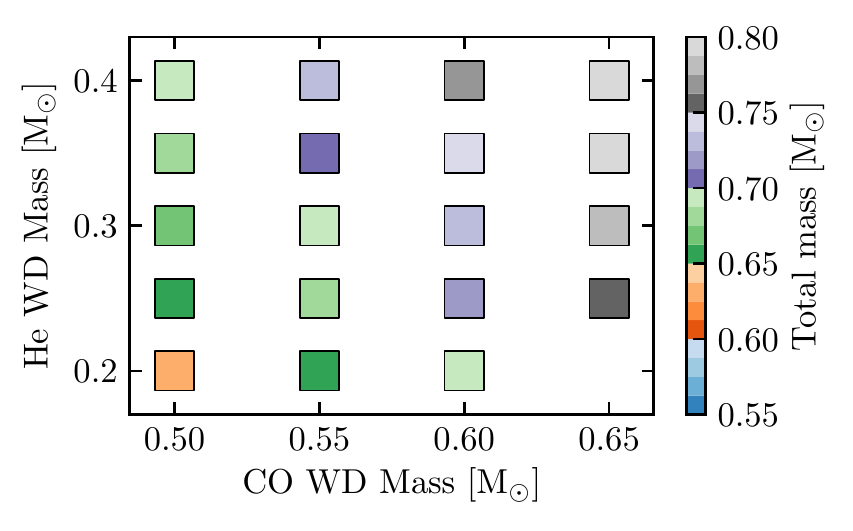}{\columnwidth}{(b) Wind with $\dot{M} = f \dot{M}_{\rm max}$ using $f = 0.001$.}
  \caption{Total mass at the end of the \RCrB\ phase when models
    reached $\Teff = \unit[10^4]{K}$.}
  \label{fig:grid-masses}
\end{figure}

The likely direct decedents of the \RCrB\ stars are the extreme He
(EHe) stars \citep[e.g.,][]{Pandey2001, Jeffery2008c}, in particular
the portion of the population that lies relatively close to the
Eddington luminosity.%
\footnote{There are also lower luminosity EHe stars which may be the
  result of He+He WD mergers \citep{Zhang2012a, Jeffery2017b}.}
The mass-luminosity relationship for EHe stars \citep{Saio1988b,
  Jeffery1988} provides an opportunity to probe their masses, which
for some objects may be aided by improved distances from \textit{Gaia}
\citep{GaiaCollaboration2016}.  The presence of radial pulsations in
these objects provides additional information and \citet{Saio1988a}
and \citet{Jeffery2001a} use spectroscopic and pulsational methods to
measure masses.  In a number of systems, multiple methods are in
agreement, leading to EHe star masses $\approx \unit[0.8-0.95]{\Msun}$
(though with significant uncertainties).  These masses are in 
tension with the lower masses of the models shown in
Figure~\ref{fig:grid-masses}. This may indicate that the amount
  of mass lost is overestimated in these models.

Alternatively, even with this mass loss, an object with
  $\gtrsim \unit[0.7]{\Msun}$ could indicate a
  CO WD primary $\gtrsim \unit[0.7]{\Msun}$, since the material that is shed
  is material from the He WD secondary.
However, these higher mass CO WD systems may be
near the lower limit for the occurrence of a detonation in the He
accretion stream during the merger itself \citep{Guillochon2010},
which can then trigger the detonation of the CO core.  (For a summary
of this double detonation scenario in the context of Type Ia SNe, see
\citealt{Shen2018b}).  Detonations of CO WDs in the mass range
$\unit[0.7-0.85]{\Msun}$ would not produce the \nickel[56] necessary
to power a Type Ia SN, but would primarily synthesize intermediate
mass elements \citep{Polin2019}.  This would still likely result in
the destruction of the system (and hence not the formation of an
\RCrB\ star).  Therefore, it may be that mergers with CO primaries
$\gtrsim \unit[0.7]{\Msun}$ and He-rich secondaries are
catastrophically destroyed during the merger process.
This picture could also fit with the suggestion by \citet{Staff2018} that
He buffer layers are required to reproduce the \oxygen[18]/\oxygen[16]
ratio, since CO WDs up to around $\unit[0.7]{\Msun}$ maintain
significant surface He layers \citep[e.g.,][]{Zenati2019}.

The relative duration of the EHe and \RCrB\ phases is
  particularly sensitive to the adopted mass loss rates.  Less mass
  loss implies a longer \RCrB\ lifetime and more core growth.  But
  then a higher core mass implies a shorter EHe lifetime, as the
  residual He layer is more rapidly exhausted at higher luminosity
  \citep{Saio1988b}.
  
\begin{figure*}
  \centering
  \includegraphics[width=\textwidth]{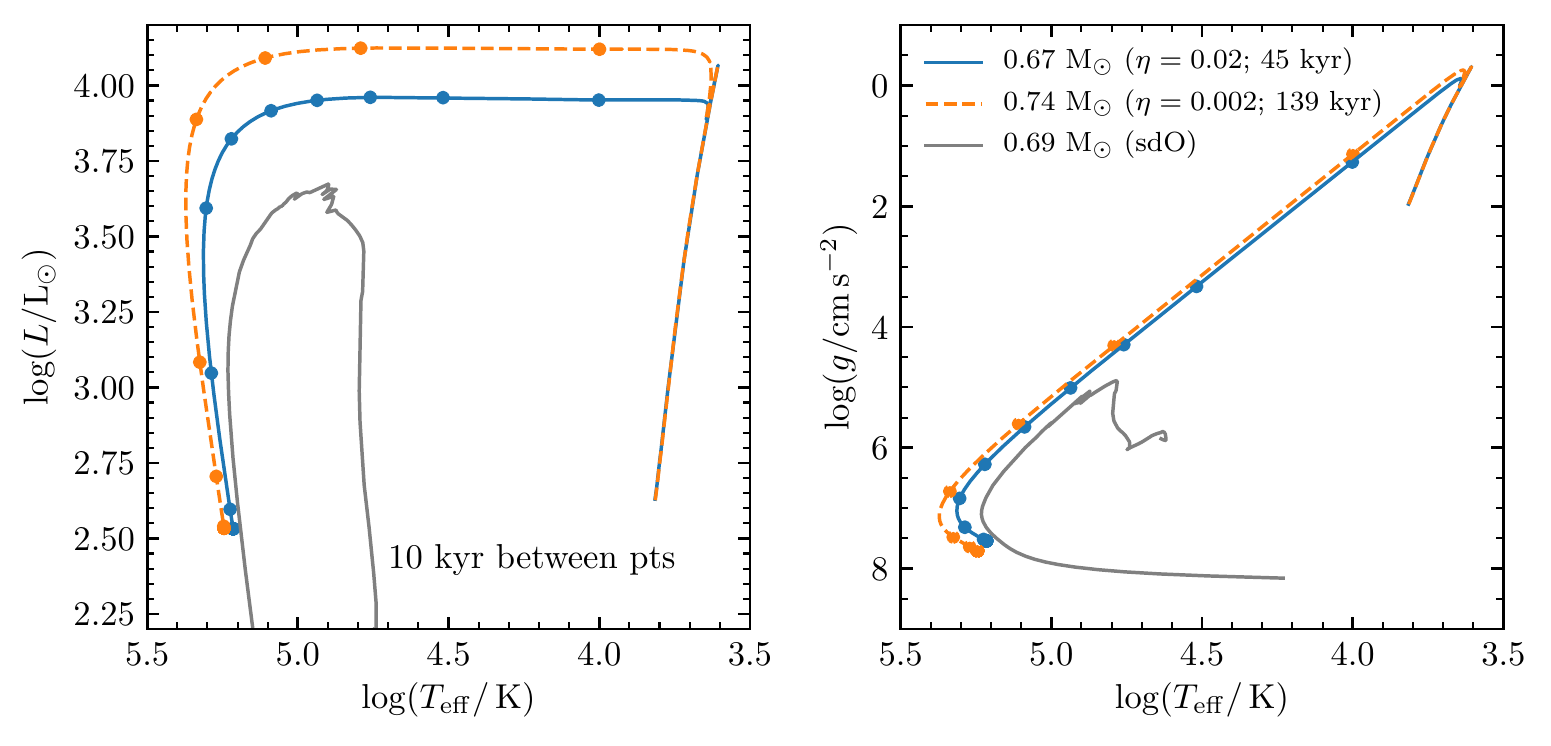}
  \caption{HR diagram (left panel) and Kiel diagram (right panel) for
    two models (colored lines) using \citet{Bloecker1995a} mass loss
    rates with different $\eta$.  The approximate time in the \RCrB\
    phase is indicated in the right legend.  Beginning when the models
    reach $\Teff = \unit[10^4]{K}$, dots along the evolutionary tracks
    are spaced in 10 kyr intervals.  The grey track shows a He + He WD
    merger model \citep[M07 from][]{Schwab2018} that has a similar
    final mass as the other models but never had a giant phase.}
  \label{fig:late-tracks}
\end{figure*}

  Figure~\ref{fig:late-tracks} shows the HR and Kiel diagrams for
  models using \citet{Bloecker1995a} winds with $\eta = 0.002$ and
  $\eta = 0.02$.  The model with the lower mass loss rate spends
  approximately three times as long in a \RCrB\ phase and becomes
  $\unit[0.07]{\Msun}$ more massive.  The dots, spaced each 10 kyr
  along the evolutionary tracks, show that this more massive remnant
  evolves blueward approximately twice as fast.  Thus, these two
  choices for the mass loss (which give final remnant masses within
  $\approx 10$ \%) are different by a factor of $\approx 6$ in the
  ratio of the predicted \RCrB-to-EHe lifetimes.
  
  The models continue to move blueward throughout the EHe phase and
  evolve to higher surface gravity where they appear as sdO stars. For
  reference, Figure~\ref{fig:late-tracks} also shows an sdO track of a
  He WD + He WD merger model from \citet{Schwab2018}.  As the models
  approach their maximum effective temperature, they appear as O(He)
  stars \citep[e.g.,][]{Rauch2008, Reindl2014a}, before finally moving
  onto the WD cooling track.  Though we do not pursue it here, there
  is potential to constrain and refine the models by exploiting the
  connections between \RCrB\ populations and their blue descendants.

\subsection{Implications for single WDs}

Eventually, the remnants of these double WD mergers will evolve into
single WDs.  The detection of single WDs with relatively short cooling
ages but kinematics consistent with ages much longer than the
single-star evolutionary timescale at their mass probe the
contribution of double WD mergers to the single WD population
\citep[e.g.,][]{Wegg2012}.  The hot DQ WDs \citep{Dufour2008} have
been speculated to show the kinematic signatures associated with
mergers \citep{Dunlap2015} as have some other massive DQs
\citep{Cheng2019}.

Immediately after the merger, the total mass of the remnant is
  approximately the sum of the component masses, with only
  $\sim \unit[10^{-3}]{\Msun}$ being ejected during the merger itself
  \citep[e.g.,][]{LorenAguilar2009}.  The mass lost during the
  subsequent evolution to a single WD then seems likely to be the
  dominant effect in setting the final mass.
  Different post-merger evolutionary pathways imprint
  themselves on the mapping from the total mass distribution of merging 
  WDs to the mass distribution of single WDs coming from merged WDs.
  
  Double He WD mergers evolve through a hot subdwarf phase
  \citep[e.g.,][]{Zhang2012a} that does not result in significant wind
  mass loss.  Some angular momentum must be shed in order for the
  remnant to reach a He core burning configuration, but this can be
  achieved while losing only $\sim \unit[10^{-2}]{\Msun}$
  \citep{Gourgouliatos2006, Schwab2018}.
  
  As discussed earlier, He WD + CO WD mergers that undergo an \RCrB\
  phase are predicted to lose $\sim \unit[10^{-1}]{\Msun}$, with
  initially more massive systems preferentially losing more mass.
  Moreover, if systems with higher mass CO WD primaries
  ($\gtrsim \unit[0.7]{\Msun}$) are destroyed during the merger
  process as a result of double detonations, this would further remove
  systems with higher total masses from the pool that will leave
  behind single WDs.  Together, these effects might lead to an
  enhancement of single WDs formed by double WD mergers around the
  masses of the EHe stars and a relative scarcity immediately above that.

  At yet higher masses ($\approx \unit[1.1]{\Msun}$),
  double CO WD mergers would begin to leave single
  WD remnants again.  These more massive merger remnants may also
  experience mass loss and hence a reduction in their total masses.
  However, since they do not set up long-lived, shell-powered burning
  giant structures \citep[e.g.,][]{Nomoto1985}, the mass loss may be
  qualitatively different and less extreme than the \RCrB\ stars.

\section{Summary \& Conclusions}
\label{sec:conclusion}

We use the \MESA\ stellar evolution code to construct evolutionary
models of stars that reach an \RCrB-like phase.  We incorporated
opacities from \AESOPUS\ \citep{Marigo2009} appropriate for the cool,
H-deficient, CNO-enhanced photospheres of these stars.  We use these
to discuss some of the model variations expected as a result of
variations in metallicity, envelope composition, and the modeling of
convection.

We construct models via the evolution of He stars
\citep[reproducing][]{Weiss1987b}.  We also construct two types of
models of the remnants of double WD mergers: one is a \MESA\
realization of the end state of the merger calculation of
\citet{Schwab2012}; the other engineers models that have a cold CO
core and high entropy He envelope, schematically reproducing the
post-merger structure.  We emphasize that models originating from
double WD merger scenarios have a thermal reconfiguration phase that
can last up to $\approx \unit[1]{kyr}$ post merger.  Some galactic
objects should be in this phase and we suggest they could be
distinguished by their lower luminosities, secular brightening, and by
the fact that their circumstellar environments should not yet have
accumulated significant dusty shells from mass loss during the \RCrB\
phase.

We illustrate, in agreement with the results of past work
\citep{Zhang2014b, Lauer2018}, that \RCrB\ models that include mass
loss based on AGB prescriptions like that of \citet{Bloecker1995a}
typically leave the \RCrB\ phase with total masses
$\approx \unit[0.7]{\Msun}$.  When the mass loss rates scale with $L$,
the steep core mass-luminosity relationship for He giants implies
convergent evolution in mass, where the initially lower mass CO WDs
grow through He shell burning but higher mass ones do not.  This
implies that the descendants of most \RCrB\ stars should have a
relatively narrow range in mass ($\approx \unit[0.1]{\Msun}$),
substantially narrower than the range in total mass of the systems
that form them.  Moreover, if double detonations
\citep[e.g.,][]{Guillochon2010, Shen2018b} are common in mergers involving
CO WDs with masses $\gtrsim \unit[0.7]{\Msun}$, then this would remove
the possibility of forming \RCrB\ stars with CO cores above this mass.
If this is true, such a limit should be reflected in the masses and
luminosities of extreme He stars and the masses of the single WDs that they eventually become.

Connecting populations of double WD systems (whether observed or from
population synthesis) to the observed numbers of \RCrB\ stars requires
accurate lifetime estimates.  In turn, accurate mass loss
prescriptions are required for accurate lifetime estimates; our \RCrB\ model
lifetimes vary by a factor of a few, depending on the assumed prescription.
As also noted by \citet{Zhang2014b}, more theoretical work that unifies the
periodic dust formation that is the \RCrB\ phenomenon itself with the
long-term average mass loss is desirable.  In the future, this would
enable models to move beyond the application of existing AGB mass loss
rates to these H-deficient stars.

\acknowledgements

This work has had a long gestation and taken many forms and directions along the way.
We acknowledge stimulating workshops at Sky House and KITP.  We thank
Lars Bildsten, Eliot Quataert, Jared Brooks, Ken Shen,
Aaron Dotter, Simon Jeffery, and Nicole Reindl
for helpful conversations.  We thank the organizers and
participants in the 2018 Hydrogen-Deficient Stars conference in Armagh
and the 2019 Hot Subdwarfs and Related Objects conference in Hendaye
for simulating meetings.
We thank the referee for a constructive report.
Support for this work was provided by NASA through Hubble Fellowship
grant \# HST-HF2-51382.001-A awarded by the Space Telescope Science
Institute, which is operated by the Association of Universities for
Research in Astronomy, Inc., for NASA, under contract NAS5-26555.
This work made use of the Hyades computing resource at UC Santa Cruz supported by NSF AST-1229745
and the Extreme Science and Engineering Discovery Environment (XSEDE), which is
supported by NSF ACI-1548562 (allocation TG-AST180050).

\software{\MESA\ \citep{Paxton2011, Paxton2013, Paxton2015, Paxton2018, Paxton2019} r11701 \citep{MESA_r11701},
  \texttt{Python} (available from \href{https://www.python.org}{python.org}),
  \texttt{matplotlib} \citep{matplotlib},
  \texttt{NumPy} \citep{numpy},
  \texttt{py\_mesa\_reader} \citep{pmr},
  \texttt{MesaScript} \citep{MesaScript}
}

\newpage
\bibliography{RCB-grid}

\end{document}